\documentclass[preprint2]{proto}
\usepackage{times}
\usepackage{psfig}
\newcommand{\refs}{\par\noindent\hangindent=1pc\hangafter=1}
\newcommand{\oversim}[2]{\protect{\mbox{\lower0.5ex\vbox{%
   \baselineskip=0pt\lineskip=0.2ex
   \ialign{$\mathsurround=0pt #1\hfil##\hfil$\crcr#2\crcr\sim\crcr}}}}} 
\newcommand{\simgreat}{\mbox{$\,\mathrel{\mathpalette\oversim>}\,$}} % >~ sign
\newcommand{\simless} {\mbox{$\,\mathrel{\mathpalette\oversim<}\,$}} % <~ sign
\newcommand{\cc}{cm$^{-3}$}
\newcommand{\kms}{km s$^{-1}$}

\begin{document}

\title{\textbf{\LARGE The Fragmentation of Cores and the Initial
    Binary Population}}

\author {\textbf{\large Simon P Goodwin}} 
\affil{\small\em University of Sheffield}  
\author {\textbf{\large  Pavel Kroupa}}
\affil{\small\em Rheinische Fredrich-Wilhelms-Universit\"at Bonn}
\author {\textbf{\large  Alyssa Goodman}}
\affil{\small\em Harvard-Smithsonian Center for Astrophysics} 
\author {\textbf{\large Andreas Burkert}}
\affil{\small\em Ludwig-Maximillian-Universit\"at Munich}

\begin{abstract}
\baselineskip = 11pt
\leftskip = 0.65in
\rightskip = 0.65in
\parindent=1pc {\small Almost all young stars are found in multiple
systems.  This suggests that protostellar cores almost always fragment
into multiple objects. The observed properties of multiple systems
such as their separation distribution and mass ratios provide strong
constraints on star formation theories.  We review the observed 
properties of young and old multiple systems
and find that the multiplicity of stars changes.  Such an evolution is
probably due to (a) the dynamical decay of small-$N$ systems 
and/or (b) the destruction of multiple systems within dense clusters.
We review simulations of the fragmentation of rotating and turbulent 
molecular cores.  Such models almost always produce multiple systems,
however the properties of those systems do not match observations at
all well.  Magnetic fields appear to supress fragmentation, prehaps
suggesting that they are not dynamically important in the formation of
multiple systems.  We finish by discussing possible reasons why 
theory fails to match observation, and the future prospects for 
this field.\\~\\~\\~}

%\end{list}
\end{abstract}  

%%%%%%%%%%%%%%%%%%%%%%%%%%%%%%%%%%%%%%%%%%%%%%%%%%%%%%%%%%%%%%%%%%%%%%%%%%%%
\section{\textbf{INTRODUCTION.}} 
%%%%%%%%%%%%%%%%%%%%%%%%%%%%%%%%%%%%%%%%%%%%%%%%%%%%%%%%%%%%%%%%%%%%%%%%%%%%

Correctly predicting the properties of young  multiple systems is one
of the most challenging tests of any theory of star formation.  In this
chapter we discuss the current understanding of how dense prestellar
cores fragment into multiple stars including brown dwarfs (by 
`stars' we generally mean both stars {\em and} brown dwarfs).  

Firstly, we will discuss the important observed properties of both
young and old binary systems and the differences between them.  Then
we will describe the possible origins of the differences between the
young and old systems and hopefully convince the reader that 
almost all stars must form in multiple systems.  This initial
binary population must form from the fragmentation of star-forming
dense molecular cores and so we discuss the observed properties of
these cores that may influence their ability to form multiple
systems.  We will then review the current models of core
fragmentation, with an emphasis on turbulence as the mechanism that
promotes fragmentation.  Finally we will examine why theory currently
fails to correctly predict binary properties.

%%%%%%%%%%%%%%%%%%%%%%%%%%%%%%%%%%%%%%%%%%%%%%%%%%%%%%%%%%%%%%%%%%%%%%%%%%%%
\section{\textbf{THE PROPERTIES OF MULTIPLE SYSTEMS.}} 
%%%%%%%%%%%%%%%%%%%%%%%%%%%%%%%%%%%%%%%%%%%%%%%%%%%%%%%%%%%%%%%%%%%%%%%%%%%%

There has been an extensive study of binary  properties over the  past
two decades with the modern study often marked as beginning  with the
detailed survey by {\em Duquennoy and Mayor} (1991, hereafter  DM91).
Multiple systems in the field are by far  the best studied due to the
availability of local samples whose completeness is easier to
estimate, and the properties of the field provide the benchmark
against which younger samples are  measured.

\bigskip
\noindent
\textbf{2.1 Multiple Systems in the field.}
\bigskip

{\em 2.1.1 Multiplicity fraction.} The fraction of field stars in multiple
systems is found to be high and increases with
increasing primary mass.  Many different measures are used to
quantify multiplicity which can become very confusing.  Most important
is the 'multiplicity frequency' $f_{\rm mult} = $
(B+T+Q+...)/(S+B+T+Q+...)  where S, B, T and Q are the numbers of
single, binary, triple and quadruple systems respectively, thus
$f_{\rm mult}$ represents the probability that any system is a
multiple system (see {\em Reipurth and Zinnecker}, 1993).

The raw value of the field G-dwarf multiplicity frequency found by
DM91 is $0.49$, when corrected for incompleteness this rises to
$0.58$.  However, recent studies using Hipparcos data have shown that
it may even be higher than this ({\em Quist and Lindegren}, 2000; {\em
S\"oderhjelm}, 2000).

It should be noted that the brown dwarf (BD) binary fraction has
generally been considered to be much lower than that of stars at
$\sim 0.1 - 0.2$ ({\em Bouy et al.}, 2003; {\em Close et al.}, 2003; 
{\em Gizis et al.}, 2003; {\em Mart\'in et al.}, 2003).  However, recent
studies have suggested that the BD multiplicity frequency may be
significantly higher, possibly exceeding 0.5 ({\em Pinfield et al.},
2003; {\em Maxted and Jeffries}, 2005), provided most BDs reside in
very tight BD--BD pairs.

\begin{figure}[t]
\centerline{\psfig{figure=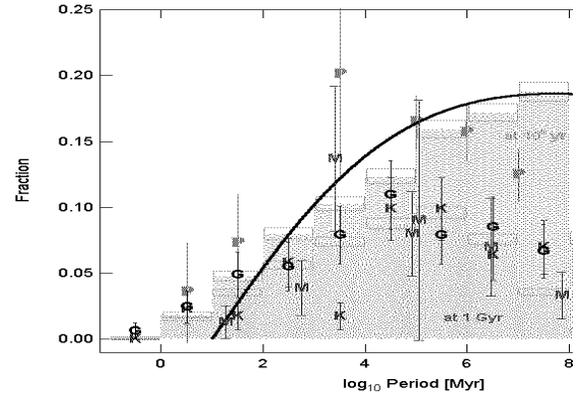,height=7.5cm,width=8.0cm,angle=0}}
 \caption{\small The {\bf period distribution} function.  Letters show
 the observed fraction of field G, K and M-stars and PMS stars (P).  The
solid curve shows the model initial period distribution  (see 
Eqn.~\ref{eq:prim_per_distr}).  The light histogram is the initial
 binary population in the simulations of {\em Kroupa} (1995b) which
 evolves through dynamical interactions in a cluster into a field-like
 distribution shown by the heavy histogram.}
\label{fig:period_distr}
 \end{figure}

{\em 2.1.2 Separation distribution.} The binary separation distribution is
very wide and flat, usually modelled as a log-normal with mean $\sim
30$~AU and variance $\sigma_{\rm log {d}} \sim 1.5$ (DM91 for
G-dwarfs).  This is illustrated in Fig.~\ref{fig:period_distr} where the
field period distribution is compared to that of young stars (note
that $a^3/P^2 = m_{\rm sys}$, where $a$ is in~AU, $P$ is in years and $m_{\rm sys}$ is the system mass in $M_\odot$).

A similar distribution is found for M-dwarfs by {\em Fischer and
Marcy} (1992), and generally seems to hold for all stars, although the
maximum separations do appear to decrease somewhat, but not
substantially, for stars with decreasing mass ({\em Close et al.},
2003).  Very low-mass stars (VLMSs) and BDs seem strongly biased
towards very close companions with semi-major axes $a\le a_{\rm max}
\approx 15$~AU ({\em Close et al.},  2003; {\em Gizis et al.}, 2003; 
{\em Pinfield et al.}, 2003; {\em Maxted and Jeffries}, 2005) in contrast
to those of stars that have $a_{\rm max}\simgreat 100$~AU (DM91; 
{\em Fischer and Marcy}, 1992; {\em Mayor et al.}, 1992).  It is this unusual
separation distribution which may have led to the underestimate of the
BD multiplicity fraction. The much smaller $a_{\rm max}$ for VLMSs and
BDs compared to the other stars cannot be a result of disruption in a cluster
environment but must be due to the inherent physics of their formation
({\em Kroupa et al.}, 2003).

\begin{figure}[t]
\centerline{\psfig{figure=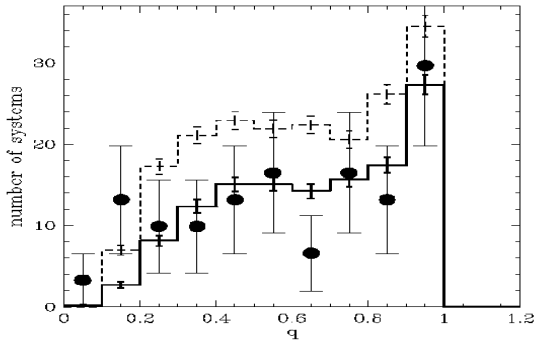,height=7.5cm,width=8.0cm,angle=0}}
\caption{\small The mass-ratio distribution of all late-type primary
stars. The solid dots are observational data by {\em Reid and Gizis}
(1997), while the expected initial distribution is shown as 
the dashed histogram. In
a typical star cluster it evolves to the solid histogram which
reproduces the observed data quite well (from {\em Kroupa et al.}, 
2003).}  \label{fig:massratio_distr}
\end{figure}

{\em 2.1.3 Mass ratio distribution.} DM91 found that Galactic-field systems
with a G-dwarf primary have a mass-ratio distribution biased towards
small values such that it does not follow the stellar IMF which would
predict a far larger number of companions with masses $m_2\simless
0.3\,M_\odot$ ({\em Kroupa}, 1995b). For short-period binaries, the
mass-ratio distribution is biased towards similar-mass pairs ({\em
Mazeh et al.}, 1992). Integrating over all periods, for 
a sample of nearby systems with primary
masses in the range $0.1 \simless m_1/M_\odot \simless 1$, {\em Reid
and Gizis} (1997) find the mass-ratio distribution to be approximately
flat and consistent with the IMF (Fig.~\ref{fig:massratio_distr}). 

{\em 2.1.4 Eccentricity distribution.} Binary systems have a thermalised
eccentricity distribution (Eqn.~\ref{eq:ecc_distr} below) for periods
$P\simgreat 10^{3-4}$~d, with tidally circularised binaries dominating
at low separations (DM91; {\em Fischer and Marcy}, 1992).

{\em 2.1.5 Higher-order systems.} DM91 find the uncorrected ratio of systems
  of different multiplicity to be S:B:T:Q = $1.28:1:0.175:0.05$ (see
  also {\em Tokovinin and Smekhov}, 2002), suggesting that roughly 20\%
  of multiple systems are high-order systems.  Concerning the origin 
of high-order multiple systems in the Galactic field, we note that 
many, and perhaps most, of these may be
the remnants of star clusters ({\em Goodwin and Kroupa}, 2005).

\bigskip
\noindent
\textbf{2.2 Pre-Main Sequence Multiple Systems.}
\bigskip

The properties of pre-main sequence (PMS) multiple systems are much
harder to determine than those in the field. We refer
the reader to the chapter by {\em Duch\^ene et al.} for a detailed
review of the observations of PMS multiple systems and the inherent
problems.

Probably the most important difference between the PMS and field
populations is that young stars have a significantly higher {\bf
  multiplicity fraction} than the field (see the chapter by 
{\em Duch\^ene et al.}; also see Fig.~\ref{fig:period_distr}).

The {\bf separation distribution} of PMS stars also appears
different to that in the field with an over-abundance of binaries with
separations of a few hundred~AU ({\em Mathieu}, 1994;  {\em Patience et
al.}, 2002; Fig.~\ref{fig:period_distr}).  More specifically, the
binary frequency in the separation range $\sim 100 - 1000$~AU is a
factor of $\sim 2$ higher than in the field ({\em Mathieu}, 1994; {\em
Patience et al.}, 2002; {\em Duch\^ene et al.}, 2004).  Extrapolating
this increase across the whole separation range implies that $f_{\rm
mult}$ for PMS stars could be as high as 100\%.  (It appears that in
Taurus the binary frequency {\em is} $\sim 100$\% for stars
$>0.3M_\odot$, {\em Leinert et al.}, 1993; {\em K\"ohler and Leinert},
1998).

The {\bf mass-ratio distribution} of PMS stars is similar to the field
population. A detailed comparison is not yet possible because low-mass
companions to pre-main sequence primaries are very difficult to
observe as the available results depend mostly on direct imaging or
speckle interferometry, while for main-sequence systems
radial-velocity surveys have been done over decades (DM91).  Thus,
using near-infrared speckle interferometry observations to obtain
resolved JHK-photometry for the components of~58 young binary systems,
{\em Woitas et al.} (2001) found that the mass-ratio distribution is
flat for mass ratios $q\ge0.2$ which is consistent with random pairing
from the IMF, i.e. fragmentation processes rather than
common-accretion (Fig.~\ref{fig:massratio_distr}).

The {\bf eccentricity distribution} of PMS stars also is similar to the
field, with a thermalised distribution except at low separations where
tidal circularisation has occurred rapidly (e.g., {\em Kroupa}, 1995b;
{\em White and Ghez}, 2001). Finally, the ages of components in
young multiple systems appear to be very similar ({\em White and Ghez},
2001).

It is currently unclear what the proportion of higher-order 
multiples is in young systems (see the chapter by {\em Duch\^ene 
et al.}).  We will revisit this question in the next section.

\bigskip
\noindent
\textbf{2.3 The evolution of binary properties.}
\bigskip

The observations described in Sections~2.1 and~2.2 clearly show that 
at least the binary fraction and separation distributions evolve 
significantly between young stellar populations and the field.  

Indeed, binary properties are seen to change even within populations
in star forming regions.  The binary fraction is found to vary
between embedded and (older) non-embedded sources in both Taurus and
$\rho$~Oph ({\em Duch\^ene et al.}, 2004; {\em Haisch et al.}, 2004).
Also, the mass ratio distributions of massive stars appear to depend on
the age of the cluster with those in young clusters being consistent
with random sampling from the IMF and those in dynamically evolved
populations favouring equal-mass companions (Section~2.4). 

\begin{figure}[t]
% \epsscale{0.9}
% \plotone{fevol_ht.ps}
\centerline{\psfig{figure=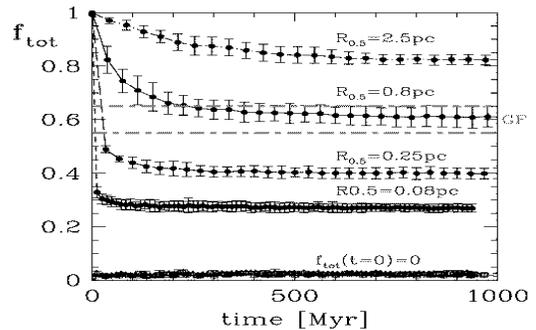,height=7.5cm,width=8.0cm,angle=0}}
 \caption{\small Evolution of the total binary fraction as a function
of time for four cluster models with decreasing half-mass radii
$R_{0.5}$ from top to bottom.  Each cluster initially contains 
200 late-type binaries (with periods, eccentricities and mass-ratios 
consistent with the initial distribution functions described in 
Section~2.4) except the bottom simulation which contains {\em no} 
binaries.  The area marked `GF' shows the observed field binary 
fraction (DM91).  The cluster with $R_{0.5} = 0.8$~pc evolves into a
field-like distribution.  With no initial binaries capture is unable
to produce a significant number of binaries.  From {\em Kroupa} (1995a).} 
\label{fig:fevol}
\end{figure}

The evolution of binary properties has been ascribed to two
mechanisms.  {\it Firstly}, the rapid dynamical decay of young
small-$N$ clusters within cluster cores (e.g., {\em Reipurth and Clarke}, 2001;
{\em Sterzik and Durisen}, 1998, 2003; {\em Durisen et al.}, 2001; {\em
Hubber and Whitworth}, 2005; {\em Goodwin and Kroupa}, 2005; {\em
Umbreit et al.}, 2005), and {\it secondly}, the dynamical destruction 
of multiples by interactions
in a clustered environment can modify an initial PMS-like
distribution into a field-like distribution ({\em Kroupa}, 1995a, b;
{\em Kroupa et al.}, 2003; Figs.~\ref{fig:period_distr},
\ref{fig:fevol}).

{\em 2.3.1 Small-$N$ decay.} Multiple systems containing $N \geq 3$ stars
are unstable to dynamical decay unless they form in a
strongly hierarchical configuration (stability criteria for $N>2$
systems are provided by {\em Eggleton and Kiseleva}, 1995).
Generally, a triple system is unstable to decay with a half-life of
\begin{equation}
t_{\rm decay} = 14 \left( \frac{R}{{\rm AU}} \right)^{3/2} \left(
\frac{M_{\rm stars}}{M_\odot} \right)^{-1/2} \,\,\,\, {\rm yrs}
\end{equation}
where $R$ is the size of the system, and $M_{\rm stars}$ is the mass
of the components ({\em Anosova}, 1986).  The decay time for $R=250$~AU
and $M_{\rm stars} = 1 M_\odot$ is $\sim 55$~kyr which is of order the
duration of the embedded phases of young stars, thus ejections should
mainly occur during the main Class 0 accretion phase of PMS objects
(e.g., {\em Reipurth}, 2000).  Indeed, one such early dynamical 
decay appears to have been observed
by {\em G\'omez et al.} (2006), and this is probably the process at
work to reduce the binary fraction between the embedded and
non-embedded stars seen by {\em Duch\^ene et al.} (2004) and 
{\em Haisch et al.} (2004).  These early dynamical processes cause
embedded protostars to be ejected from their natal envelopes, possibly
causing abrupt transitions of objects from class 0/I to class II/III 
({\em Reipurth}, 2000; {\em Goodwin and Kroupa}, 2005).

Significant numbers of small-$N$ decays will dilute any initial 
high multiplicity
fraction very rapidly to a small binary fraction for the whole
population (e.g., $N=5$ systems would lead to a population with a
binary fraction $f=1/5$ within $<10^5$~yr). The observed high binary
fraction in about 1~Myr old populations thus suggests that the
formation of $N>2$ systems is the exception rather than the rule ({\em
Goodwin and Kroupa}, 2005).  Ejections would occur mostly during the 
very early Class~0 stage such that ejected embryos later appear as 
free-floating single very low-mass stars and BDs ({\em
Reipurth and Clarke}, 2001). However, the small ratio of the number of
BDs per star, $\approx 0.25$ ({\em Munech et al.},
2002; {\em Kroupa et al.}, 2003; {\em Kroupa and Bouvier}, 2003b; {\em
  Luhman}, 2004), again suggests this not to be a very common process
even if {\em all} BDs form from ejections.

Ejections have two main consequences: a significant reduction in the
semi-major axis of the remaining stars ({\em Anosova}, 1986; {\em
  Reipurth}, 2000; 
{\em Umbreit et al.}, 2005); and the preferential ejection of the
lowest mass component ({\em Anosova}, 1986; {\em Sterzik and Durisen},
2003).  The early ejection of the lowest-mass component forms the
basis of the embryo ejection scenario of BD formation ({\em
Reipurth and Clarke}, 2001; {\em Bate et al.}, 2002).

The $N$-body statistics of the decay of small-$N$ systems has been
studied by a number of authors ({\em Anosova}, 1986; {\em Sterzik and
Durisen}, 1998, 2003; {\em Durisen et al.}, 2001; {\em Goodwin  et
  al.}, 2005; {\em Hubber and Whitworth}, 2005).  However, only {\em
Umbreit et al.} (2005) have attempted to include the effects of
accretion on the $N$-body dynamics which appear to have a significant
effect - especially on the degree of hardening of the binary after
ejection.  {\em Goodwin et al.} (2004a, b) and {\em Delgado Donate et al.}
(2004a, b) have simulated ensembles of cores including the full
hydrodynamics of star formation, however proper statistical
conclusions about the effects of ejections are difficult to draw due
to the different numbers of stars forming in each ensemble (which
there is no way of controlling a priori), and the smaller number of
ensembles that may be run in a fully hydrodynamic context.  However,
some conclusions appear from these and other studies ({\em Whitworth
et al.}, 1995; {\em Bate and Bonnell}, 1997; {\em Bate et al.}, 2003;
{\em Delgado Donate  et al.}, 2003).  Firstly, that early ejections
are very effective at hardening the remaining stars (c.f. {\em Umbreit
et al.}, 2005).  Secondly, this early hardening tends to push the mass
ratios of close binaries towards unity.  This occurs as the low-mass
component has a higher specific angular momentum than the primary and so
is more able to accrete mass from the high angular momentum
circumstellar material (see {\em Whitworth et al.}, 1995; {\em Bate and
Bonnell}, 1997).  However, {\em Ochi et al.} (2005) find in detailed 2D
simulations of accretion onto binaries that the gas accretes mainly
onto the primary due to shocks removing angular momentum.

One significant caveat to the previous discussion is that the gradual 
formation (over $\sim 0.1$~Myr) of stars
allows far more stable triples and higher-order multiples to form than
expected observationally, which are stable for at least 10~Myr ({\em Delgado
Donate et al.}, 2004a, b).  Indeed, simulations that form a large number
of stars often form very  hierarchical higher-order multiples (often
quadruples and quintuples formed when even larger systems decay) which
are not observed ({\em Delgado Donate et al.}, 2003, 2004a, b; {\em Goodwin et
al.}, 2004a, b).  Such systems would probably be destroyed during the
cluster destruction phase (see below), but not dilute the binary
fraction on very short timescales.

{\em 2.3.2 Dynamical destruction in clusters.}  In the 
highly-clustered environments in which most stars are thought
to form (e.g., {\em Lada and Lada}, 2003) dynamical interactions will be
common and may disrupt many initially binary systems.

Binaries can be sub-divided into three dynamical groups: (i)
the wide, or soft, binaries, (ii) the dynamically active binaries, and
(iii) the tight or hard binaries.

{\it Wide binaries} have orbital velocities much smaller than the
velocity dispersion, $\sigma$, in a cluster and are easily
disrupted. This is best seen by a gedanken experiment, where we
construct a reduced-mass particle (a test particle orbiting in a fixed
potential with total mass, eccentricity and orbital period equal to
that of the binary in question) in a heat bath of field particles (the
cluster stars). If the orbital velocity of the test particle is
smaller than the typical velocities of the field particles ($v_{\rm
orb}\ll \sigma$), then the test particle will gain kinetic energy by
encounters, i.e. by the principle of energy equipartition, until it's
orbital velocity surpasses the binding energy of the binary. The
binary consequently gets disrupted. Energy conservation requires the
heat bath to cool; cluster cooling has been seen in $N$-body
computations by {\em Kroupa et al.}, (1999), but the
effect is not significant for cluster dynamics. The general effect of
this process is that binaries with weak binding energies are disrupted
(i.e., binaries with long periods and/or small mass-ratios).

{\it Hard binaries}, on the other hand, can be represented by a
reduced-mass binary in which the test particle has $v_{\rm orb}\gg
\sigma$, so that energy equipartition leads to a reduction of $v_{\rm
orb}$ and to an increase of $\sigma$ (cluster heating).  This
increases the binding energy of the binary which heats up further
($v_{\rm orb}$ increases as the test particle falls towards the
potential minimum).  This run-away process only stops because either
the binary merges when it is so tight that the constituent stars touch
(forming a blue straggler), or because the hardened binary receives a
re-coil expelling it from the cluster, or the hardening binary evolves
to a cross-section so small that the binary becomes essentially
unresolved in further interactions.

{\em Heggie} (1975) and {\em Hills} (1975) studied the details of
these processes and formulated the {\it Heggie--Hills law} of binary
evolution in clusters: ``soft binaries soften and hard binaries
harden''. An important implication of this law is that hard binaries
can absorb the entire binding energy of a cluster and drive the
evolution of the core of a massive cluster.

Not accessible to analytical work are {\it active
  binaries} with intermediate binding energies. Only
full-scale $N$-body computations can treat the dynamics of the
interactions accurately (e.g., {\em Heggie et al.}, 1996).  Such
binaries couple efficiently to the cluster, and efficiently exchange
energy with it. The binary--binary and binary--single-star
interactions form complex resonances and short-lived higher-order
configurations that decay by expelling typically the least massive
member. Active binaries are thus quite efficient in exchanging
partners, but more work needs to be done in order to quantify the
exchange rates for typical Galactic star-forming clusters.

The dynamical interactions between binaries or single stars will
continue to alter the binary properties of the population as long as
it remains relatively dense (i.e. until the cluster dissolves or
expands significantly after residual gas removal).  In a series of
papers ({\em Kroupa}, 1995a, b; {\em Kroupa et al.}, 2001) it has been
shown that a population initially composed entirely of binary systems
with a PMS binary separation distribution can evolve into the
field-like distribution through dynamical encounters, as is
exemplified by the evolution of the mass-ratio distribution
(Fig.~\ref{fig:massratio_distr}).

Importantly, however, dynamical interactions in a cluster cannot form
a significant number of binaries from an initially single star population, 
{\em or} widen an initially narrow separation
distribution ({\em Kroupa and Burkert}, 2001). Also, the clusters
within which most Galactic-field stars form do not sufficiently 
harden an initially wide separation
distribution to be consistent with the number of tight binaries. Such
clusters that would lead to significant hardening would disrupt too
many of the wide binaries diluting the Galactic-field binary-star
population.

This indicates that the observed broad period distribution (Fig.~1) is
already imprinted at the time of binary formation.
The existing $N$-body simulations of young clusters tend to assume a
relatively well-mixed and relaxed initial distribution of
stars. However, real young clusters tend to be lumpy and unrelaxed
which alters the binary-binary interaction rate (e.g., {\em Goodwin and
Whitworth}, 2004), but is not expected to change the general outcome (a
smaller-$N$ system with a smaller radius is {\it dynamically
equivalent} to a larger-$N$ system with a larger radius, ({\em Kroupa},
1995a).

{\em 2.3.3 The effect of dynamics on the binary population.} Small-$N$ 
decay will act to modify the {\em birth binary population}
(ie. that produced by star formation) by reducing the overall 
binary fraction and by hardening the 
remaining binaries on a timescale of $<10^5$ yrs.  Small-$N$ decay
cannot occur very often as it would produce too many single PMS 
stars and too many hard binaries (and quite possibly too many BDs).
However, it is unclear if small-$N$ decay is rare because cores
usually form only binaries and only sometimes triples, or because higher
order multiples are formed in stable, hierarchical systems.

The {\em initial binary population} (ie. the population after
small-$N$ decay and internal energy re-distribution processes,
``eigenevolution''-see below, has modified it) is further changed by
dynamical interactions within a cluster on a timescale of a~Myr for
typical Galactic star-forming clusters ({\em Kroupa}, 1995a, b; {\em
Adams and Myers}, 2001).  Encounters will destroy soft and active
binaries leaving mostly the hard binary population unchanged.  Crucially
it cannot produce more binaries in any significant numbers.

Thus both of the processes that act to modify the birth binary
population into the field binary population {\em reduce} the binary
fraction.  We are led to the conclusion that the birth binary fraction
must be higher than that of the field.  This conclusions agrees well
with observations.

{\em Goodwin and Kroupa} (2005) argue that the birth binary fraction must
be almost unity for all stars, and that the low binary fraction amongst
M-dwarfs is due to the preferential destruction of low-mass binaries.
{\em Lada} (2006), however, argues that most stars form as single stars,
as most M~dwarfs are single and most stars are M~dwarfs.  The crucial
issue here is how many initial M~dwarf binaries decay?  We note that in a
model that assumes that stars are born with a 60~\% binary
fraction with companion masses selected randomly from the IMF
and without dynamical dissolution of the binaries leads to
a population with the observed binary
fraction-spectral type relationship ({\em Kroupa et al.}, 1993, their
Fig.~11).  In models that assume a 100~\% initial binary fraction,
processing through an `average' cluster ({\em Lada and Lada}, 2003) also
converts the binary population into the observed field population.
Observations of the binarity of a very young (embedded?) M~star population
in an average cluster (as opposed to Taurus which is atypical) are
required to fully resolve this issue.  However, for higher-mass stars, it
seems almost certain that the initial binary fraction must be {\em
significantly} higher than the field binary fraction due to their current
relatively high binary fraction.  It is worth noting that the observed
population of high-order multiples may be formed from the final
handful of stars that remain at the end of cluster dissolution which
tend to be in high-order heiarchical systems (see {\em Goodwin and
  Kroupa}, 2005).

Thus, whilst more work needs to be done, especially on the effects of
the cluster density and on the initial distribution in star-forming
regions with very different physical properties, a picture has emerged
in which the observed high multiplicity PMS population can be modified
by secular dynamical evolution to produce the field population. 
Most (especially K-dwarfs and later) 
stars therefore form in multiple - probably binary and triple -
systems with a very wide separation distribution and
a relatively flat mass-ratio distribution.

\bigskip
\noindent
\textbf{2.4 The initial binary population.}
\bigskip

Given these results, a useful working hypothesis appears to be that
the initial binary-star properties are
invariant to star-formation conditions.  The observed differences
between binary populations result from different secular dynamical
histories of the respective populations: ie. due to different cluster masses
and densities.

{\rm Kroupa} (1995b) suggests that the field binary properties can be
understood if the {\em birth} binary population has the following
semi-empirical distribution functions:

\vspace{0.05cm} \noindent (1.) Companion masses are chosen randomly from the IMF;

\vspace{0.05cm} \noindent (2.) The distribution of periods is
independent of primary mass, and can be described with the following 
functional form,
\begin{equation}
f_{lP} = 2.5 {lP-1 \over 45 + (lP-1)^2},
\label{eq:prim_per_distr}
\end{equation}
where $dN_{lP} = N_{\rm tot}\,f_{lP}\,dlP$ is the number of binaries
with periods in the range $lP,lP+dlP$ ($lP\equiv {\rm log}_{10}P$, $P$ in days) and
$N_{\rm tot}$ is the number of single-star and binary-star systems in
the sample under consideration;

\vspace{0.05cm} \noindent (3.) The eccentricity distribution is 
thermal (all binding energies are equally occupied),
\begin{equation}
dN=f_{\rm o}\,f(e)\,de = f_{\rm o}\,2\,e\,de,
\label{eq:ecc_distr}
\end{equation}
being the number of binary systems with eccentricity in the range
$e,e+de$.

\vspace{0.05cm}

These birth distributions need to be modified for short-period 
binaries ($lP\simless 3$) through the evolution of the binding energy 
and angular momentum owing to dissipative processes within the young 
binary system termed collectively as pre-main sequence eigenevolution.  
This then gives the {\it initial} distributions which are
evident in dynamically unevolved populations (e.g., Taurus, {\em
Kroupa and Bouvier}, 2003a) and can be used as the initial binary-star
population in $N$-body modelling of stellar populations.

\begin{figure}[t]
\centerline{\psfig{figure=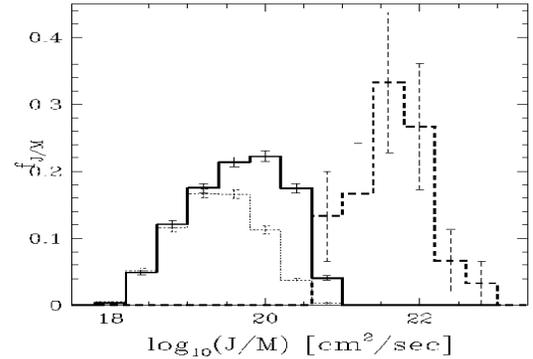,height=7.5cm,width=8.0cm,angle=0}}
 \caption{\small The observed specific angular momentum distribution
of molecular cloud cores by {\em Goodman et al.} (1993) is shown as
the rightmost dashed histogram, while the initial binary-star
population (Section~2.4) is plotted as the solid histogram. It evolves to
the dot-dashed histogram after passing through a typical star
cluster (from {\em Kroupa}, 1995b).}
\label{fig:specific_angmom}
\end{figure}

The distribution over binding energies and specific-angular momenta
can be evaluated readily given the above distribution
functions. Fig.~\ref{fig:specific_angmom} shows that the distribution
of specific angular momenta of molecular cloud cores forms a natural
extension to the distribution of specific angular momenta of the
initial binary stellar population, possibly suggesting an
evolutionary connection.

These semi-empirical distribution functions have been formulated for
late-type stars (primary mass $m_{\rm p}\simless 1\,M_\odot$) as it is
for these that we have the best observational constraints.  It is not
clear yet if they are also applicable to massive binaries.  {\em
Baines et al.}  (2006) report a very high ($f\approx 0.7\pm 0.1$)
binary fraction among Herbig Ae/Be stars with the binary fraction
increasing with increasing primary mass. Furthermore, they find that
the circumbinary discs and the companions appear to be co-planar
thereby supporting a fragmentation origin rather than collisions or
capture as the origin of massive binaries.  Most O~stars are believed
to exist as short-period binaries with $q\approx 1$ ({\em
Garc{\'{\i}a} and Mermilliod}, 2001), at least in rich clusters, while
small $q$ appear to be favoured in less substantial clusters such as
the Orion Nebula Cluster (ONC), being consistent there with random
pairing ({\em Preibisch et al.},  1999). {\em Kouwenhoven et al.}
(2005) report the A and late-type B binaries in the Scorpius OB2
association to have a mass-ratio distribution {\it not} consistent
with random pairing. The lower limit on the binary fraction is~0.52.
Perhaps the massive binaries in the ONC represent the primordial
population, whereas in rich clusters and in OB associations the
population has already dynamically evolved through hardening and
companion exchanges to that observed there. This possibility needs to
be investigated using high-precision $N$-body computations of young
star clusters.

Given such reasonably-well quantified estimates of the distribution
functions of orbital elements of the primordial binary population, the
problem remains as to how these distributions functions can be
understood theoretically as a result of the star-formation
process. {\em Fisher} (2004) notes that distribution functions similar
to the ones derived above, and in particular a wide mass-ratio
distribution, a very wide period range, and a thermal eccentricity
distribution, are obtained quite naturally from a turbulent molecular
cloud (see also {\em Burkert and Bodenheimer}, 2000).

%%%%%%%%%%%%%%%%%%%%%%%%%%%%%%%%%%%%%%%%%%%%%%%%%%%%%%%%%%%%%%%%%%%%%%%%%%%%
\section{\textbf{THE PROPERTIES OF PRESTELLAR CORES.}} 
%%%%%%%%%%%%%%%%%%%%%%%%%%%%%%%%%%%%%%%%%%%%%%%%%%%%%%%%%%%%%%%%%%%%%%%%%%%%

The gas that is just-about-ready to form stars arranges itself into
denser structures often called prestellar `cores' (e.g., {\em Myers 
and Benson}, 1983).  Often, a `typical' prestellar core is described as
having a radius $\sim 0.1$~pc, density $\simgreat10^4$~\cc, and
velocity dispersion $\sim 0.5$~\kms.  In fact though, the idea that
such cores are `typical' primarily arises from the relative ease with
which nearby, isolated dense cores, that will each form fewer than a
handful of stars, can be observed and modelled.  It is in fact likely
that accounting for the diversity in core properties is crucial to
improving the match between theory and observations of the conversion
of gas to (binary) stars.

\bigskip
\noindent
\textbf{3.1 What is a `core'?}
\bigskip

In general, observations and theory have concentrated on {\em 
isolated} and {\em coherent} prestellar cores such as those found 
in low-mass star forming regions such as Taurus (due to the relative
ease of observing such cores).  It is not yet clear if legitimate 
analogs to these cores exist within the dense concentrations of 
gas that form the clusters (e.g., {\em Goodman et al.}, 
1998).  In particular, the $\sim 0.1$~pc size of isolated cores would
result in them having multiple dynamical encounters in the dense
environment that forms `typical' clusters such as Orion (e.g., {\em
  Lada and Lada}, 2003).  

That we have not observed any 0.1~pc core analogues in dense clusters
is not surprising.  Even in very nearby clusters like NGC1333 in Perseus
(at $\sim 300$~pc), 0.1~pc is 1~arcmin, typical of single-dish
resolution for tracers like NH$_3$ and N$_2$H$^+$, which map out gas
with density $\simgreat 10^4$ (e.g., {\em Benson and Myers}, 
1989; {\em Evans}, 1999).   Thus, to find meaningful dense structures 
on scales significantly less than 1~arcmin, interferometry is 
required.   Interferometers have definitively revealed sub-structure
in the gas within clusters, but this substructure does not offer a
one-to-one gas clump-to-star match the way observations of isolated
cores do.  Instead, regions forming many stars are associated with
more dense gas than those forming fewer.  This lack of one-to-one
correspondence suggests that long-lived blobs associated with the
formation of individual cores within clusters do not exist.

Given this, is it reasonable to extend observations and simulations of
isolated cores to more typical clustered star forming `cores'?  The
answer is possibly.  Whilst large isolated cores cannot exist in
clustered star forming regions, the regions in which stars are thought
to form are far smaller than the size of a whole core.
Observationally, the size of binary systems is $<$a few hundred~AU in
agreement with theoretical expectations of the scale of fragmentation
(see Section~4.1).  Systems of this scale would be expected to
interact on timescales of $>$~Myr in a typical cluster which is
significantly longer than the star formation timescale is thought 
to be (see the chapters by {\em Di Francesco et al.} and 
{\em Ward-Thompson et al.}).  So, while the details of the first 
stages of collapse
in isolated cores are probably not applicable to most star formation,
the details of the final stages of fragmentation and star formation
occurring on few hundred~AU scales quite possibly occur in relative
isolation.  However, continued accretion onto cores may significantly
effect the evolution of the inner proto-stellar system depending on
the details and timescale of core and star formation in
clusters.

With this in mind, we will continue to review the properties of
isolated pre-stellar cores.

\bigskip
\noindent
\textbf{3.2 Rotation.}
\bigskip

Clearly, for cores to fragment some angular momentum must be present
otherwise the cores will collapse onto a single, central  point.   The
simplest source of this angular momentum is due to bulk rotation of
the core.  
It is a relatively straightforward procedure to estimate the component
of solid-body rotation present in a dense core by fitting for the
gradient in observed line-of-sight velocity over the face of a core
(e.g., {\em Goodman et al.}, 1993; {\em Barranco \& Goodman}, 
1998; {\em Caselli et al.}, 2002).   The results of this 
fitting (see Fig.~4) have
been used as input values of `initial angular momentum' in many
calculations.  While the estimates of the {\it component} of
solid-body rotation made in this way are sound, and are thus fine to
use as inputs, it is important to appreciate that cores do not really
rotate as solid-bodies ({\em Burkert and Bodenheimer}, 
2000).  When the velocity
measurements are put into the context of measurements of velocities on
larger scales, both observations ({\em Schnee et al.}, 2005) and simulations
({\em Burkert and Bodenheimer} 2000), show that the ``rotation" is often just an
artifact created by larger-scale turbulent motions.

Fig.~\ref{fig:specific_angmom}  gives a summary of the measured 
specific angular momentum for core rotation for the 29 dense 
cores from {\em Goodman et al.} (1993) which show significant
rotation.  The majority of cores included in the
{\em Goodman et al.} (1993) study are isolated, low-mass cores: one
should keep in mind that the rotational properties of smaller
fragments that may form inside those cores as true(r) precursors 
to protostars remain largely unmeasured.

\bigskip
\noindent
\textbf{3.3 Non-thermal line widths.}
\bigskip

It has been known for many years ({\em Larson}, 1981; {\em Myers},
1983; {\em Solomon et al.}, 1987; see also {\em Elmegreen and Scalo}, 
2004a and references therein) that the
line widths inside even the most quiescent of dense cores are more
than thermal.  The coldest isolated dense cores have gas temperatures
of order 10~K, and dust temperatures as low as 6~K (see the chapters
by {\em Di Francesco et al.} and {\em Ward-Thompson et al.}).
A gas temperature of 10~K implies an H$_2$ 1$\sigma$ velocity 
dispersion of only 0.2~\kms.  Observed dispersions have a 
distribution from $\sim 0.2$ to 1~\kms with a peak at $\sim 0.4$~\kms, 
never quite reaching down to the thermal value (e.g., see the
catalogue of {\em Jijina et al.}, 1999).

The origin of the non-thermal line width in dense cores is the subject
of an extensive literature, but it is fair to say that a consensus
exists that `turbulence' is responsible (see e.g., {\em Myers}, 
1983; {\em Barranco \& Goodman}, 1998; {\em Goodman et al.}, 
1998; {\em Elmegreen and Scalo}, 2004a).  Significant levels of turbulence
in cores are important for fragmentation as they may provide the
angular momentum to form multiple systems (see Section~4.2.2) and, as
mentioned above, could also be responsible for the observed rotation.

\bigskip
\noindent
\textbf{3.4 Magnetic fields.}
\bigskip

The role of magnetic fields in supporting cores against collapse 
is a subject of much debate.  Magnetic fields are not thought to be 
dynamically dominant in cores as was once thought (e.g., {\em Shu et
  al.}, 1987; or indeed in the ISM as a whole {\em Elmegreen and
  Scalo}, 2004a). However, they may be very important as the role of 
magnetic fields in the fragmentation of cores is poorly understood
(see Section~4.2.6).

Isolated cores are found to be statistically triaxial with a tendency
towards being prolate ({\em Jones et al.}, 2001; {\em Goodwin et al.},
2002).  In contrast to this, magnetic support would tend to 
produce oblate cores.  In addition these magnetically-supported oblate 
cores would tend to rotate around their short axis which is not
observed ({\em Goodman et al.}, 1993).  This is supported by
observations of the magnetic field which show that cores are not
magnetically critical ({\em Crutcher et al.}, 1999; see {\em Bourke 
and Goodman}, 2004 and references therein).

%%%%%%%%%%%%%%%%%%%%%%%%%%%%%%%%%%%%%%%%%%%%%%%%%%%%%%%%%%%%%%%%%%%%%%%%%%%%
\section{\textbf{THE FRAGMENTATION OF PRESTELLAR \\ CORES.}} 
%%%%%%%%%%%%%%%%%%%%%%%%%%%%%%%%%%%%%%%%%%%%%%%%%%%%%%%%%%%%%%%%%%%%%%%%%%%%

\bigskip
\noindent
\textbf{4.1 The Physics of Collapsing Cores.}
\bigskip

During the early stages of the collapse of a prestellar core, the rate
of compressional heating is low and the gas is able to cool
radiatively, either by molecular line emission, or, when $\rho >
10^{-19}$ g cm$^{-3}$, by thermally coupling to the dust.  The  gas is
therefore approximately isothermal (at $\sim 10$~K) with an equation
of state $P \propto \rho$.

Eventually the rate of compressional heating becomes so high (due to
the acceleration of the collapse), and the rate of radiative cooling
so low (due to the increasing column density and dust optical depth),
that the gas switches to being approximately adiabatic, with $P
\propto \rho^\gamma$ (where $\gamma = 5/3$ initially for a monatomic
gas, and then $\gamma=7/5$ above $\sim 300$~K when H$_2$ becomes
rotationally excited).

This behaviour has been studied in detail ({\em Larson}, 1969; {\em
Tohline}, 1982; {\em Masunaga et al.}, 1998; {\em Masunaga and Inutsuka},
2000) for cores in the range $1 - 10 M_\odot$ with an initial
temperature of 10~K.  These authors find that the switch between
isothermality and adiabaticity occurs at a critical density of
$\rho_{\rm crit} \sim 10^{-13}$ g cm$^{-3}$.  (See Fig.~1 in {\em
Bate} 1998 or Fig.~2 in {\em Tohline}, 2002). Thus, as contraction
proceeds and the density, $\rho$, increases the Jeans mass, $M_{\rm
J}\propto \rho^{-1/2}T^{3/2}$, decreases as long as the gas can retain
the same temperature, $T$, while it is optically thin. Once the
opacity increases such that the gas core heats up, $M_{\rm J}$
increases.  The most important result of this thermal behaviour is
therefore that there is a {\em minimum} Jeans mass that is reached at
$\sim \rho_{\rm crit}$ of order $M_{\rm min} \sim 10^{-2} M_\odot$.
This is often referred to as the opacity limit for fragmentation.

There is an even lower minimum mass that occurs during a later
isothermal phase at $\rho \sim 10^{-3}$ g cm$^{-3}$ when molecular
hydrogen dissociates at a few thousand K.  It is possible that a
further fragmentation episode can occur at these densities which may
account for some close binaries.

Fragmentation in cores is expected to occur at around $\rho_{\rm
crit}$ as at lower or higher densities the  Jeans mass increases
(although how rapidly it rises above $\rho_{\rm crit}$ does depend
sensitively on the $\gamma$ used  in the adiabatic equation of state).
Thus we expect multiple systems to be formed with a typical length
scale $R_{\rm form}$ of
\begin{equation}
R_{\rm form} < \left( \frac{3M_{\rm core}}{4\pi \rho_{\rm crit}}
\right)^{1/3} \sim 125~(M_{\rm core}/M_\odot)^{1/3}\,\,\,{\rm AU}
\end{equation}
where $M_{\rm core}$ is the mass of the core.  Interestingly, this
scale matches the observed peak in the T~Tauri separation distribution
(see also {\em Sterzik et al.}, 2003).

There is a minimum separation in this picture of $\sim 30$~AU which is
the separation of two fragments of $M_{\rm min}$ at $\rho_{\rm crit}$.
It may - or may not - be significant that this is the {\em average}
binary separation (DM91; see also {\em Sterzik et al.}, 2003).  
However, it would appear difficult to form
binaries closer than $\sim 20 - 30$~AU without some hardening
mechanism {\em or} a secondary fragmentation phase.

It should be noted that the length scales of star formation of less
than a few hundred~AU are several orders of magnitude smaller than the
thousands of~AU scales on which core properties have been observed.

\bigskip
\noindent
\textbf{4.2 Fragmentation mechanisms.}
\bigskip

In this section we examine the main mechanisms that have been proposed
to explain multiple formation.  Given the complex and highly
non-linear nature of the physics in most models, numerical simulations
are the main route by which the mechanisms for fragmentation have been
investigated.  Bulk rotation and turbulence are the two main
mechanisms that have been considered to provide the angular momentum
required for fragmentation to occur and we review the theoretical work
and simulations conducted on both of these mechanisms.  In addition,
we discuss the possible role of magnetic fields, disc fragmentation
and `secondary fragmentation'.

{\em 4.2.1 Rotational fragmentation.}  The simplest situation in which 
fragmentation may well occur is in a
spherical cloud with solid-body rotation and an isothermal equation of
state.  {\em Tohline} (1981), using semi-analytic arguments concluded
that all such clouds should fragment.  A number of simulations have
shown that such clouds do fragment if $\alpha_{\rm therm} \beta_{\rm
rot} \simless 0.12-0.15$,  where $\alpha_{\rm therm} = E_{\rm
therm}/|\Omega|$ is the initial  thermal virial ratio (where $E_{\rm
therm}$ is the thermal kinetic and $\Omega$ is the gravitational
potential energy), and $\beta_{\rm rot} = E_{\rm rot}/|\Omega|$ the
initial rotational virial ratio (where $E_{\rm rot}$ is the rotational
kinetic energy) ({\em Miyama et al.}, 1984; {\em Hachisu and Eriguchi},
1984, 1985; {\em Miyama}, 1992; see also {\em Tsuribe and Inutsuka},
1999a, b; {\em Tohline}, 2002).

{\em Boss and Bodenheimer} (1979) added an $m=2$ azimuthal density
perturbation to a standard rotating cloud (effectively creating an
elongated cloud more similar to those observed than purely spherical
clouds; see Section~3.4).  They found that with a
perturbation of amplitude $A=0.5$ the cloud fragments into a
binary system.  This simulation was repeated by {\em Burkert and
Bodenheimer} (1993) who also found that when $A=0.1$ a filament
connecting the two components of the binary fragments into several
smaller fragments.  However the  connecting filament should not
fragment as predicted by {\em Inutsuka  and Miyama} (1992) and
demonstrated by {\em Truelove et al.} (1997).  Indeed the 'Boss and
Bodenheimer test' has become a standard test for the accuracy of codes
(e.g., {\em Truelove et al.}, 1997 for adaptive mesh refinement - AMR -
and {\em Kitsonias and Whitworth}, 2002 for smoothed particle
hydrodynamics - SPH).  However, it is a rather unsatisfactory test as,
whilst the {\em Truelove et al.} simulations are generally considered
to have converged, no analytic solution to the problem exists.  An
alternative test based on the original analysis of Jeans is presented
by {\em Hubber et al.} (2006).

The simulation of rotating clouds can be made more physical by
including an adiabatic (e.g., {\em Tohline}, 1981; {\em Miyama}, 1992) or
barotropic (e.g., {\em Bonnell}, 1994; {\em Bate and Burkert}, 1997; {\em
Boss et al.}, 2000; {\em Cha and Whitworth}, 2003) equation of state
(eos).  {\em Bate and Burkert} (1997) showed that the {\em Boss and
Bodenheimer} test {\em does} produce a line of fragments with a
barotropic eos, but not if it remains isothermal.  In addition, {\em
Boss et al.} (2000) simulated a cloud with an $m=2$, $A=0.1$
perturbation using a barotropic eos {\em and} also with radiation
transport; the second case producing a binary whilst the first did
not: despite the similarity of the pressure-temperature relations.
Both of these results suggest that fragmentation is highly sensitive
to thermal inertia and radiation transport effects.

Other authors have modified the initial conditions to include effects
such as different density profiles (e.g., {\em Myhill and Kaula}, 1992; {\em
Burkert et al.}, 1997; {\em Boss}, 1996; {\em Boss and Myhill}, 1995;
{\em Burkert and Bodenheimer}, 1996; {\em Boss et al.}, 2000; {\em Boss},
1993), differential rotation (which tends to promote fragmentation:
{\em Myhill and Kaula}, 1992; {\em Boss and Myhill}, 1995; {\em Cha and
Whitworth}, 2003) and non-spherical shapes (e.g., {\em Bastien}, 1983;
{\em Bonnell and Bastien}, 1991; {\em Bonnell et al.}, 1991; {\em Nelson
and Papaloizou}, 1993; {\em Boss}, 1993; {\em Sigalotti and Klapp},
1997).  The effect increasing external pressure on the collapse of
rotating cores have been investigated  by {\em Hennebelle et al.}
(2003, 2004, 2006).

{\em 4.2.2 Turbulent fragmentation.}  Recently a picture of 
star formation as a rapid and highly dynamic
process has appeared (e.g., {\em Elmegreen}, 2000; {\em
V\'aquez-Semadeni et al.}, 2000; {\em Larson}, 2003; {\em Elmegreen 
and Scalo}, 2004b) as opposed to a
quasi-static process (e.g., {\em Shu et al.}, 1987).  In particular, the
idea of cores evolving slowly via ambipolar diffusion (e.g., {\em Basu
and Mouschovias}, 1994, 1995a, b; {\em Ciolek and Mouschovias},
1993, 1994, 1995; {\em Ciolek and Basu}, 2000) has been replaced by one
in which cores form in converging flows in a highly turbulent
molecular cloud.   This is rather good news for fragmentation, as the
main effects of a  quasi-static evolution are to delay fragmentation
and reduce the  angular momentum and turbulence in a core and organise
material  so that its collapse is well focused onto a central point
(see  also Section~3.2).  Simulations of core formation in a
turbulent medium suggest that cores form with significant amounts  of
turbulence.  Turbulent, rapidly formed cores also reproduce many of
the observed properties of cores (e.g., {\em Burkert and Bodenheimer},
2000; {\em Ballesteros-Paredes et al.}, 2003; {\em Jappsen and Klessen},
2004) and have a mass spectrum not dissimilar to the observed core
mass spectrum (e.g., {\em Padoan and Nordlund}, 2002, 2004; {\em Klessen et
al.}, 2005).

Simulations of the effects of turbulence in cores focus on two
different regimes: high-velocity ({\em Bate et al.}, 2002, 2003; {\em Bate and
Bonnell}, 2005; {\em Delgado Donate et al.}, 2004a, b), and low-velocity ({\em
Goodwin et al.}, 2004a, b).  Also see {\em Fisher} (2004) for a
semi-analytic  approach to multiple formation with turbulence.  The
level of  turbulence is usually quantified as a turbulent virial ratio
$\alpha_{\rm turb} =  E_{\rm turb}/|\Omega|$, where $E_{\rm turb}$ is
the kinetic energy in turbulent motions and $|\Omega|$ is the gravitational
potential energy (note - not any rotational property).
Highly turbulent simulations
focus on $\alpha_{\rm turb} = 1$ in $50 M_\odot$ ({\em Bate et al.},
2002, 2003; {\em Bate and Bonnell}, 2005) and $5 M_\odot$ ({\em Delgado
Donate et al.}, 2004a, b) cores.  Simulations of slightly turbulent cores
range between $\alpha_{\rm turb} = 0 - 0.25$ in $5.4  M_\odot$ cores
({\em Goodwin et al.}, 2004a, b).  In all of these simulations turbulent
motions are modelled using a Gaussian divergence-free random velocity
field $P(k) \propto k^{-n}$ where $n$ is usually taken to be $4$ to
match observations of cores for which $n=3-4$ provides a good fit to
the Larson relations ({\em Burkert and Bodenheimer}, 2000).  It should
be noted that the random chaotic effects introduced by
variations in  the initial turbulent velocity field can be very
important.  Therefore a statistical approach is desirable utilising
large ensembles of simulations (e.g., {\em Larson}, 2002).

These simulations of turbulence are different to those of turbulence
in molecular clouds which concentrate on the
formation of dense cores and massive stars (e.g., {\em Klessen 
and Burkert}, 2000).  This is largely due to
computational limitations which do not allow the resolution of the
opacity limit for fragmentation in the large-scale context of giant
molecular clouds.  A mass resolution of $\sim 10^{-2} M_\odot$ is
required to resolve the opacity limit for fragmentation in SPH.  In
AMR the problem is even worse as the Jeans length continues to fall
(albeit more slowly) after the minimum Jeans mass is reached and codes
must resolve few~AU scales to capture the lowest-mass fragments.

Even very low levels of turbulence ($\alpha_{\rm turb} \sim 0.025$)
are enough to allow cores to fragment: that is, for most cores in  an
ensemble to form more than one star ({\em Goodwin et al.}, 2004b).  As
the level of turbulence is increased, the average number of stars that
form in a core increases ({\em Goodwin et al.}, 2004b).  It has been
suggested that  approximately one star forms per initial Jeans mass:
$\sim 1 M_\odot$ for these initial temperatures and densities
(cf. {\em Bate et al.}, 2003; {\em Delgado Donate et al.}, 2004a).  This
seems to hold in highly-turbulent cores, however the number of stars
forming falls with decreasing turbulence ({\em Goodwin et al.}, 2004b)
and so this at best probably only represents a (statistical)
asymptotic behaviour.

In highly-turbulent cores, the supersonic turbulent velocity field
creates a number of condensations in shocked, converging regions which
become Jeans unstable and collapse (see {\em Bate et al.}, 2003; {\em
Delgado Donate et al.}, 2004a, b).  However, it is unclear if this mode of
fragmentation is realistic in small (certainly $<$ a few $M_\odot$)
cores as the observed levels of non-thermal motions rule-out
significant highly supersonic turbulence in these cores.

\begin{figure*}
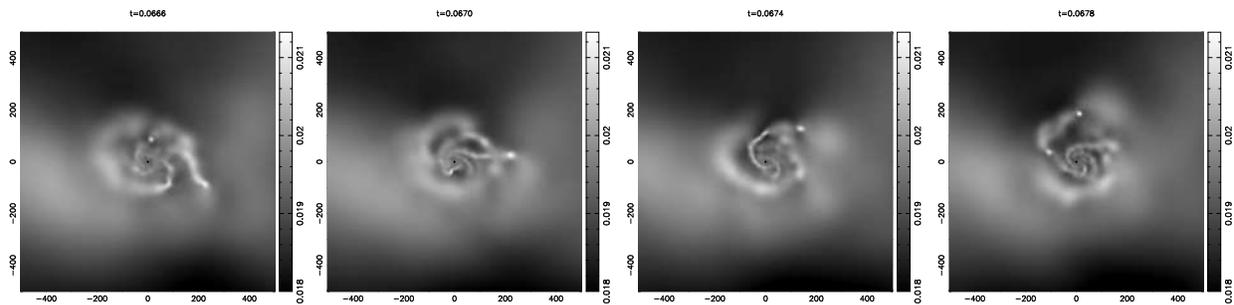

\centerline{\psfig{figure=Goodwinfig5a.ps,height=3.9cm,width=3.9cm,angle=270}\,\,\,\psfig{figure=Goodwinfig5b.ps,height=3.9cm,width=3.9cm,angle=270}\,\,\,\,\psfig{figure=Goodwinfig5c.ps,height=3.9cm,width=3.9cm,angle=270}\,\,\,\,\psfig{figure=Goodwinfig5d.ps,height=3.9cm,width=3.9cm,angle=270}}
 \caption{\small The evolution of a CAR in a turbulent molecular core
 from {\em Goodwin et al.} (2004a).  The boxes are 1000~AU on a side
 and the timescale runs from 66.6 to 67.8~kyr after the start of the
 simulation in steps of 400~yrs.  The grey-scale bar
 gives the column density in g cm$^{-2}$.  The spiral features can
 become self-gravitating if their density exceeds $\sim 10^{-12}$ g
 cm$^{-3}$ as their Jeans length falls to $\sim 20$~AU which can allow
 collapse without being shredded.  This does not always occur, in the
 first two panels a dense knot can be seen being shredded and
 accreted onto the central object.  Very similar evolution is also
 seen in AMR simulations ({\em Gawryszczak et al.}, 2006).
 }\label{fig:sims}
\end{figure*}

In Fig.~\ref{fig:sims} we show the formation of a fragment in a
mildly-turbulent $5.4M_\odot$ core with $\alpha_{\rm turb} = 0.05$
based on observations of the isolated core L1544 (from {\em Goodwin 
et al.}, 2004a).  Fragmentation occurs in a `disc-like' mode in
circumstellar accretion regions (we avoid the use of `disc' to
describe these regions as they are not rotationally supported
structures) (CARs) which form around the first star.
CARs are highly unstable structures as
there is non-uniform (in space, time and angular momentum) inflow onto
them.  Complex spiral instabilities form in the CAR due to this
inhomogeneous infall of material.  We note that these instabilities
are seen in both SPH {\em and} AMR simulations of the same situation
({\em Gawryszczak et al.}, 2006; see also Section~4.4).  Fragmentation
occurs if the density in spiral waves becomes high enough that the
Jeans length falls to  the typical width of a spiral wave and the
collapse time falls to a  low enough fraction of the local rotation
period that it may escape  shredding by differential rotation.  In
these simulations, fragmentation occurs for some (but not all) regions
that exceed $\sim 10^{-12}$ g cm$^{-3}$ in density (equating to a
Jeans length of $\sim 20$~AU) beyond $\sim 50-100$~AU from the central
star.  We note that the highly unstable nature of CARs makes usual
applications of instability criteria such as the Toomre Q-parameter
impossible.

Such a mode of fragmentation is highly sensitive to the equation of
state that has been adopted.  It has been found that the number of
fragments that form increases if $\gamma$ is changed from $5/3$ to
$7/5$ ({\em Goodwin et al.}, in prep.).  This is due to the sensitivity
of the Jeans length with density and so to the ease with which
fragmentation can occur in CARs.

The process of fragmentation in CARs is highly chaotic, relying as it
does on a certain degree of `luck' in being able to reach a
high-enough density and avoiding shredding whilst collapsing.  Thus it
is no surprise that anywhere between 1 and 12 stars form in each core
depending entirely on the details of the initial turbulent velocity
field ({\em Goodwin et al.}, 2004a, b).

In summary, it is found that turbulent cores generally fragment into
several stars: approximately one per initial Jeans mass ($\sim 1
M_\odot$) in the core.  The number of stars that form increases with
increasing turbulence and is also highly sensitive to the details of
the turbulent velocity field.  However, only relatively high-mass
cores ($> 5 M_\odot$) have been investigated in turbulent simulations
so far.  The effect of turbulence in lower-mass cores must be
investigated, as lower-mass cores appear to dominate the core mass
function ({\em Motte et al.}, 1998; {\em Testi and Sargent}, 1998;
{\em Motte et al.}, 2001).

{\em 4.2.3 Disc fragmentation.}  Disc fragmentation is a 
mechanism by which low-mass stars and BDs 
may be formed.  In the dense environments of clusters close
encounters between stars can disturb the circumstellar discs promoting
instabilities which can lead to the fragmentation of otherwise stable
discs.  (Note that this is rather different to the turbulent disc-like
scenario described above as these proto-planetary discs are much less
massive than CARs and are also stable, rotationally supported discs as
opposed to CARs).

In a series of papers, {\em Boffin et al.} (1998) and {\em Watkins}
(1998a,b) found that most star-disc interactions will lead to
gravitational  instabilities which form new low-mass companions.
These simulations generally considered massive discs where $M_{\rm
star} = M_{\rm disc} = 0.5 M_\odot$.  {\em Bate et al.} (2003) find
that star-disc encounters play an important role in forming binaries
and also truncating discs.   Star-disc encounters are also thought to
play an important role in redistributing angular momentum in
proto-planetary discs even if they do not cause further fragmentation
({\em Larson}, 2002; {\em Pfalzner}, 2004; {\em Pfalzner et al.}, 2005).

Star-disc encounters probably play a role in star formation, and may
lead to to the formation of BD (or even planetary) mass
companions ({\em Thies et al.}, 2005).  However, they are
probably not a significant contributor to the primordial stellar
binary population.  This is due to the requirement that the encounters
occur early in the star formation process - during the Class 0 phase
when the disc mass is still very large compared to the stellar mass -
a phase which lasts for only $\sim 10^5$ yrs, leaving only a small
time for encounters to occur.  However, the role of disc fragmentation
in planet formation may well be important.

{\em 4.2.4 The role of magnetic fields.}  The treatments 
of collapse and fragmentation discussed above do  not
include magnetic fields.  The new picture of rapid, turbulence-driven
star formation combined with the lack of observational evidence for
magnetically critical  cores suggests that magnetic fields are not
dynamically dominant.  In addition, fragmentation is expected to occur
at densities $\simgreat 10^{-13}$ g cm$^{-3}$,  densities at which the
magnetic field is expected to be decoupled from the gas due to the
extremely low fractional ionisation (see {\em Tohline}, 2002).
However, possibly one of the main reasons for neglecting magnetic
fields is the difficulty in including them in SPH simulations.
(although this is  improving, see esp. {\em Hosking and Whitworth},
2004a, b; {\em Price and Monaghan}, 2004a, b).  Magnetic fields {\em are}
clearly present in (many) cores, even if they are not dynamically
dominant, and their effects may be very  important.

Grid-based simulations which include magnetic fields in rotating
clouds show that fragmentation can occur in these clouds, although
magnetic fields appear to have a tendency to suppress fragmentation
(e.g., {\em Hosking and Whitworth}, 2004b; {\em Machida et al.},  2005b),
although {\em Boss} (2002;2004) claims the opposite.  {\em Sigalotti
and  Klapp} (2000) find binary and higher-order multiple formation in
slowly rotating $\sim 1 M_\odot$ clouds which includes a model for
ambipolar diffusion.

Possibly the most extensive investigation of the effects of magnetic
fields on fragmentation has been made by {\em Machida et al.}
(2005a,b).   They find that fragmentation occurs in $\sim 50\%$ of
their  rotating, magnetised clouds when either the rotation is
relatively  high or magnetic field strength relatively low.  In
particular  fragmentation always occurs in magnetised clouds if
$\beta_{\rm rot}  > 0.05$, but it almost never occurs below this limit
(see Fig.~10  from {\em Machida et al.}, 2005b). Indeed, {\em Burkert 
and Balsara} (2001) conclude that once magnetic fields are strong
enough to affect the dynamical evolution they will also efficiently
suppress fragmentation which means that magnetic fields cannot be important
as we know that fragmentation {\em must} occur.

\bigskip
\noindent
\textbf{4.3 `Secondary' fragmentation.}
\bigskip

As briefly mentioned in Section~4.1, there is a second isothermal
phase in the evolution of gas towards stellar densities.  This occurs
at a temperature of $\sim 2000$~K and a density of $\sim 10^{-3}$ g
cm$^{-3}$ when molecular hydrogen dissociates into atomic hydrogen.
This phase occurs in the hydrostatic protostar when its radius is
$\sim 1$~AU and - if fragmentation can occur at this stage - it may
explain very close binaries.

Both {\em Boss} (1989) and {\em Bonnell and Bate} (1994) simulated the
collapse of a rotating hydrostatic first object to high densities.
They found that fragmentation can occur in axisymmetric instabilities
or a ring formed by a centrifugal bounce.  However, {\em Bate} (1998)
found that spiral instabilities remove angular momentum and suppress
further fragmentation.  Recent 2D simulations by {\em Saigo and
  Tomisaka} (2006) suggest that the angular momentum of the first core
is a crucial factor in determining if fragmentation will occur during
the second collapse.

Thus it is unclear if a secondary fragmentation phase occurs.
However,  we suggest that such a phase could well be responsible  for
the apparently high incidence of very close BD-BD binary  systems
({\em Pinfield et al.}, 2003; {\em Maxted and Jeffries}, 2005) as the
evolution of BD-mass hydrostatic objects occurs  on a longer
timescale than in stellar-mass objects.  This possibility is being
investigated by SG without any firm conclusions as yet.

\bigskip
\noindent
\textbf{4.4 Simulations vs. Observations}
\bigskip

A summary of the simulations to date suggests that collapsing cores are
easily able to fragment.  However, no detailed model is currently able
to correctly predict all of the observed binary properties.

A successful model of star formation must produce multiple systems
which generally have only 2 or 3 stars with a wide range of
separations from $<<1$~AU to a peak at $\sim 100 - 200$~AU.  At all
separations, most stars must usually have quite different masses, but
avoiding BDs within at least 5~AU of the primary (the BD desert).  

Possibly the most significant problem at the moment, is that
simulations seem to form {\em too many} single stars (see {\em Bouvier et
al.}, 2001; {\em Duch\^ene et al.}, 2004; {\em Goodwin and Kroupa},
2005).  As described in Section~2.3.1 systems with $N \geq 3$ are generally
unstable and decay by ejecting their lowest mass member and hardening
the remaining multiple.  The ejection of members of small-$N$ multiple
systems dilutes the multiplicity of stars, as ejected stars tend to be
single.  Thus, many ejections will result in a far lower multiplicity
fraction than is observed in young star forming regions.  {\em Goodwin
and  Kroupa} (2005) suggest that the observed multiplicity frequencies
can  be explained if roughly half of cores form 2 stars, and half form
3  stars.  However, these numbers are far lower than are usually found
in core fragmentation simulations.

The inclusion of magnetic fields produces the opposite problem that
too few binaries are produced.  {\em Machida et al.} (2005b) find the
fragmentation does not occur in rotating, magnetised clouds when
$\beta_{\rm rot} \simless 0.04$ - a higher level of rotation than is
observed in many cores.  This problem becomes
especially accuse when we consider that much of the observed rotation
in cores could well be due to turbulent motions rather than a bulk
rotation.

\begin{figure*}
 \epsscale{1.5} \plotone{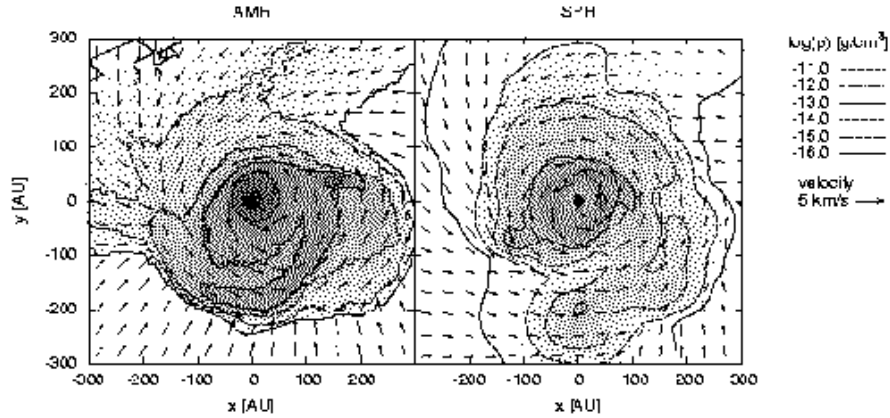}
%\centerline{\psfig{figure=fragment.ps,height=6.0cm,width=10.0cm,angle=0}}
 \caption{\small Comparison of an AMR (left) and a SPH (right)
   simulation of a collapsing, turbulent $5.4 M_\odot$ core showing
   the first `disc' fragmentation episode from {\em Gawryszczak et
   al.} (2006).  The bound fragments can be seen  at (0,125)~AU in the
   AMR simulation (left) and (-200,0)~AU in the SPH simulation
   (right).  The scale and orientation of both views are identical.
   }
\label{fig:amrsph}
\end{figure*}

A related  problem is posed by the existence of a significant number
($\sim 20\%$ of field G-dwarfs; DM91) of close, unequal mass binary
systems.  It appears difficult to form stars much closer together than
$\sim 30$~AU.  In order to obtain hard binary
systems, a further hardening mechanism is required.  In many
simulations of turbulent star formation, this hardening mechanism is
provided by the ejection of low-mass components (e.g., {\em Bate et
al.}, 2003; {\em Delgado Donate et al.}, 2004a, b; {\em Goodwin et al.},
2004a, b; {\em Umbreit et al.}, 2005; {\em Hubber and Whitworth},
2005; see {\em Ochi et al.}, 2005 for a caveat).
However, ejections appear to occur rapidly, during the main accretion
phase,  producing many close, equal-mass binary systems (see above) in
contradiction to observations of a relatively flat mass ratio
distribution ({\em Mazeh et al.}, 1992). {\em White and Ghez} (2001) do
find many roughly equal-mass PMS binaries $<100$~au in Taurus, however
there is no trend to more equal mass ratios at very low separations.
{\em Fisher et al.} (2005) do find a bias towards equal-mass
binaries among local field spectroscopic binaries (with separations
$\simless 1$~AU).  

In general, simulations that produce a small number of stars
consistent with limits on ejection do not produce hard binaries.
These binaries are difficult to form in significant numbers through
later dynamical interactions in a clustered environment, which tend to
disrupt wide binaries but not harden them.  But
simulations that produce many stars tend to form too many close,
equal-mass binaries and very high-order multiple systems (many
quadruples and quintuples) and also dilute the multiplicity fraction
too much through ejections.

\bigskip
\noindent
\textbf{4.5 Numerical issues}
\bigskip

As already discussed in this volume by {\em Klein et al.}, there is 
some debate about the ability of simulations to
correctly resolve fragmentation.  Simulations of star formation are
usually conducted using SPH as opposed to AMR schemes due to the
Lagrangian nature of SPH (see {\em Gawryszczak et al.}, 2006 for
details).

No numerical scheme is perfect, and both SPH and AMR have their
advantages and disadvantages.  It is worth noting that a  recent study
by {\em Gawryszczak et al.} (2006) has shown that AMR and SPH converge
when simulating the collapse of a slightly turbulent core.  
Fig.~\ref{fig:amrsph} shows a snapshot of both the SPH and AMR 
simulations at roughly 76~kyr from the start of the simulation showing 
that in both numerical schemes a highly unstable CAR forms that fragments in
both simulations.  {\em Gawryszczak et al.} found that AMR is
significantly more computationally intensive than SPH for an identical
simulation.  This result is not surprising as when simulating
gravitational collapse the Lagrangian nature of SPH should prove
highly efficient.  The agreement of two very different methods when
applied to the same physical situation increases our confidence that
the results are not dominated but numerical effects.

Both SPH and AMR suffer from problems with artificial angular momentum
transport.  In AMR, rotation in a poorly resolved Cartesian grid is
likely to transport angular momentum outwards.  In SPH, the use of
artificial viscosity to reduce particle inter-penetration in shocks
produces an inward transport of angular momentum with rotation.  These
problems are probably responsible for the different CAR (disc) sizes
between SPH and AMR seen by {\em Gawryszczak et al.} (2006; see also
Fig.~\ref{fig:amrsph}).

It should be noted that {\em Hubber et al.} (2006) have shown that SPH
supresses artificial fragmentation rather than promoting it which
suggests that the current generation of SPH simulations could {\em
  underestimate} the number of fragments that form.

We would conclude that the computational situation is not perfect and
many problems with both SPH and AMR remain.  However, we think that
the conflict between simulation and observation is more probably due
to missing and/or incorrect physics, rather than any fundamental
numerical difficulties.

The computational situation is made more problematic by the need to
perform ensembles of simulations to get the statistical properties of
multiple systems.  Even in cases where the result for any {\em single}
set of parameters might be expected to converge there is a large
parameter space to cover, and small differences in initial conditions
may make a significant difference to the result.  More realistically,
situations where fragmentation occurs due to non-linear instabilities,
or with turbulence (when the details of the velocity field vary)
require large ensembles even for the {\em same} region of parameter
space.  No matter how detailed or correct any single simulation is, it
can only  be a snapshot of the outcome of a particular initial
configuration.  This vastly increases the computational effort
required to model fragmentation and star formation.

\bigskip
\noindent
\textbf{4.6 Missing physics}
\bigskip

One of the greatest problems facing the simulation of core
fragmentation is to correctly model the thermal physics of cores.  It
is not possible in the foreseeable future that we will be able to
conduct hydrodynamical simulations that include a proper treatment of
radiation transport as the computational expense is just too great.
However, as shown by {\em Boss et al.} (2000) thermal effects can be
very important.  The use of the barotropic equation of state is at
best a first approximation, and it is clear that varying the adiabatic
exponent can have significant effects on fragmentation ({\em Goodwin et
al.}, in prep.).  In particular, the barotopic equation of state is
based on simulations that have used (necessarily) simplistic,
spherically symmetric assumptions.  It is not clear to what extent
these can be applied to highly inhomogeneous cores including discs and
local density peaks.  Improvements are being made, including using a
flux-limited diffusion approximation ({\em Whitehouse and Bate}, 2004)
and an approximation based on the local potential as a guide to
optical depth ({\em Stamatellos et al.}, in prep.).  However a problem
remains with any approximation in that, while it should obviously
match the fully detailed radiative transfer simulations of simple
situations, it is  not clear if it is correct in more complex
situations.

Most simulations (especially SPH simulations) do not include the
effects of magnetic fields.  Yet those simulations which do include
magnetic fields seem to suggest that fragmentation is suppressed.  This
could well be a very important conclusion given that non-magnetic
models seem to over-produce stars in cores.  However, the very
efficient suppression of fragmentation by magnetic fields may rule-out
the importance of magnetic fields in the fragmentation process as we
know that cores {\em must} fragment.

Given that fragmentation appears to
occur in disc-like structures, the proper treatment of these is vital.
Both AMR and SPH have problems with the artificial transport of
angular momentum (see {\em Gawryszczak et al.}, 2006) which will effect
their ability to correctly model discs.  Discs are also a situation in
which magnetic fields may play an important role.

Finally, very few simulations attempt to model the effects of feedback
from stars as jets or through their radiation field (for some first
attempts to deal with these problems see {\em Stamatellos et al.},
2005; {\em  Dale et al.}, 2005).  In particular, {\em Stamatellos et
al.} (2005) find that the inclusion of jets may inhibit fragmentation
by decreasing the inflow rate onto the disc and forcing that inflow to
occur away from the poles.

Many of the physical situations which may result in fragmentation are
rather complex and often chaotic (turbulence being the most obvious
example).  Such situations will not produce any single, unique answer.
Indeed, given the variety of multiple systems such a situation would
not be expected.  However, this does require that a statistical
approach be taken when performing simulations.  This vastly increases
the computational effort required, as any `single' region of an
already huge parameter space will require an ensemble of simulations
to investigate it.

%%%%%%%%%%%%%%%%%%%%%%%%%%%%%%%%%%%%%%%%%%%%%%%%%%%%%%%%%%%%%%%%%%%%%%%%%%%%
\section{\textbf{CONCLUSIONS}}
%%%%%%%%%%%%%%%%%%%%%%%%%%%%%%%%%%%%%%%%%%%%%%%%%%%%%%%%%%%%%%%%%%%%%%%%%%%%

Almost all young stars are found in multiple systems with a very wide
separation distribution and a fairly flat mass ratio distribution.
Thus prestellar cores must fragment into multiple stars and/or BDs 
with these properties.

The dynamical decay of small-$N$ systems would rapidly produce a large
single-star pre-main sequence population if large numbers of unstable 
systems form  with $N>2$ or $3$.  This decay would also result in large
numbers of very close binary systems.  Neither of these are observed,
leading to the conclusion that cores must usually form only 2 to 4
stars in hierarchical systems (for $N>2$).

Simulations show that most cores which contain some angular  momentum
- either in bulk rotation, or in turbulence - are able to fragment
into multiple objects.  However, these simulations have been
unsuccessful in matching their results to the observed young multiple
population.  In particular, the distributions of separations and  mass
ratios from simulations tend not to fit well.

The future is somewhat rosier, however.  The inclusion of more
detailed physics and more realistic initial conditions may well  yield
better fits to observations.  

%%%%%%%%%%%%%%%%%%%%%%%%%%%%%%%%%%%%%%%%%%%%%%%%%%%%%%%%%%%%%%%%%%%%%%%%%%%%
\vspace{0.1cm}
\textbf{ Acknowledgements.} SPG is supported by a UK Astrophysical
Fluids Facility (UKAFF) fellowship.

\bigskip

%%%%%%%%%%%%%%%%%%%%%%%%%%%%%%%%%%%%%%%%%%%%%%%%%%%%%%%%%%%%%%%%%%%%%%%%%%%%
\centerline\textbf{REFERENCES}
%%%%%%%%%%%%%%%%%%%%%%%%%%%%%%%%%%%%%%%%%%%%%%%%%%%%%%%%%%%%%%%%%%%%%%%%%%%%
\bigskip
\parskip=0pt {\small
\baselineskip=11pt

\refs Adams F. C. and Myers  P. C. (2001), {\em \apj, 553}, 744-753.

\refs Anosova Zh. P. (1986) {\em \apss, 124}, 217-241.

\refs Baines D., Oudmaijer R., Porter J., and Pozzo M. (2006), {\em \mnras}, 
      in press (astro-ph/0512534)

\refs Ballesteros-Paredes J., Klessen R. S., and V\'azquez-Semadeni
E. (2003) {\em \apj, 592}, 188-202.

\refs Barranco J. A. and Goodman A. A. (1998) {\em \apj, 504}, 207-222.

\refs Bastien P. (1983) {\em \aap, 119}, 109-116.

\refs Basu S. and Mouschovias T. Ch. (1994) {\em \apj 432}, 720-741.

\refs Basu S. and Mouschovias T. Ch. (1995a) {\em \apj, 452}, 386-400.

\refs Basu S. and Mouschovias T. Ch. (1995b) {\em \apj, 453}, 271-283.

\refs Bate M. R. (1998) {\em \apj, 508}, L95-L98.

\refs Bate M. R. and Bonnell I. A. (1997) {\em \mnras, 285}, 33-48.

\refs Bate M. R. and Bonnell I. A. (2005) {\em \mnras, 356},
1201-1221.

\refs Bate M. R. and Burkert A. (1997) {\em \mnras, 288}, 1060-1072.

\refs Bate M. R., Bonnell I. A., and Bromm V. (2002) {\em \mnras,
332}, L65-L68.

\refs Bate M. R., Bonnell I. A., and Bromm V. (2003) {\em \mnras,
339}, 577-599.

\refs Benson P. J. and Myers P. C. (1989) {\em \apjs, 71}, 89-108.

\refs Boffin H. M. J., Watkins  S. J., Bhattal, A. S., Francis  N., and
Whitworth  A. P. (1998) {\em \mnras, 300}, 1189-1204.

\refs Bonnell I. A. (1994) {\em \mnras, 269}, 837-848.

\refs Bonnell I. A. and Bastien P. (1991) {\em \apj, 374}, 610-622.

\refs Bonnell I. A. and Bate M. R. (1994) {\em \mnras, 269}, L45-L48.

\refs Bonnell I. A., Martel H., Bastien P., Arcoragi J-P., and Benz
W. (1991) {\em \apj, 377}, 553-558.

\refs Boss A. P. (1989) {\em \apj 346,} 336-349.

\refs Boss A. P. (1993) {\em \apj, 410}, 157-167.

\refs Boss A. P. (1996)  {\em \apj, 468}, 231-240.

\refs Boss A. P. (2002) {\em \apj, 568}, 743-753.

\refs Boss A. P. (2004) {\em \mnras, 350}, L57-L60. 

\refs Boss A. P. and Bodenheimer P. (1979) {\em \apj, 234} 289-295.

\refs Boss A. P. and Myhill E. A. (1995) {\em \apj, 451}, 218-224.

\refs Boss A. P., Fisher R. T., Klein R. I., and McKee C. (2000)  {\em
\apj, 528}, 325-335.

\refs Bourke T. L. and Goodman A. A. (2004) In {\em Star Formation at
  High Angular Resolution} (M. Burton et al., eds.), pp. 83-96. ASP
  Conf. Series, San Francisco.

\refs Bouvier J., Duch\^ene G., Mermilliod J.-C., and Simon T. (2001)
{\em \aap 375}, 989-998.

\refs Bouy H., Brandner W., Mart\'in E. L., Delfosse X., Allard F., and
Basri G. (2003) {\em \aj, 126}, 1526-1554.

\refs Burkert A. and Balsara D. (2001) {\em AAS, 33}, 885.

\refs Burkert A. and Bodenheimer P. (1993)  {\em \mnras, 264},
798-806.

\refs Burkert A. and Bodenheimer P. (1996)  {\em \mnras, 280},
1190-1200.

\refs Burkert A. and Bodenheimer P. (2000)  {\em \apj, 543}, 822-830.

\refs Burkert A. and Lin D. N. C. (2000) {\em \apj, 537}, 270-282. 

\refs Burkert A., Bate M. R., and Bodenheimer P. (1997)  {\em \mnras,
289}, 497-504.

\refs Caselli P., Benson P. J., Myers P. C., and Tafalla M. (2002) {\em
  \apj, 572}, 238-263.

\refs Cha S.-H. and Whitworth A. P. (2003)  {\em \mnras, 340}, 91-104.

\refs Ciolek G. E. and Basu S. (2000) {\em \apj, 529}, 925-931.

\refs Ciolek G. E. and Mouschovias T. Ch. (1993) {\em \apj, 418},
774-793.

\refs Ciolek G. E. and Mouschovias T. Ch. (1994) {\em \apj, 425},
142-160.

\refs Ciolek G. E. and Mouschovias T. Ch. (1995) {\em \apj, 454},
195-216.

\refs Close L. M., Siegler N., Freed M., and Biller B. (2003) {\em
 \apj, 587}, 407-422.

\refs Crutcher R. M., Troland T. H., Lazareff B., Paubert G., and 
Kaz\'es I. (1999) {\em \apj, 514}, L121-L124.

\refs Dale J. E., Bonnell I. A., Clarke C. J., and Bate M. R. (2005)
      {\em \mnras, 358}, 291-304.

\refs Delgado Donate E. J., Clarke C. J., and Bate M. R. (2003) {\em
  \mnras, 342}, 926-938.

\refs Delgado Donate E. J., Clarke C. J., Bate M. R., and Hodgkin,
S. T. (2004a) {\em \mnras, 351}, 617-629.

\refs Delgado Donate E. J., Clarke C. J., and Bate M. R. (2004b) {\em
  \mnras, 347}, 759-770.

\refs Duch\^ene G.,  Bouvier J., Bontemps S., Andr\'e P., and Motte
F. (2004)  {\em  \aap, 427}, 651-665.

\refs Duquennoy A. and Mayor M. (1991) {\em \aap, 248}, 485-524. (DM91).

\refs Durisen R. H., Sterzik M. F., and Pickett B. K. (2001)  {\em
  \aap, 371}, 952-962.

\refs Eggleton P. and Kiseleva L.(1995), {\em \apj}, 455), 640-645. 

\refs Elmegreen B. G. (2000)  {\em  \apj, 530}, 277-281.

\refs Elmegreen B. G. and Scalo J. (2004a), {\em \araa, 42}, 211-273.

\refs Elmegreen B. G. and Scalo J. (2004b), {\em \araa, 42}, 275-316.

\refs Evans N. J. (1999) {\em \araa, 37}, 311-362.

\refs Fischer D. A. and Marcy G. W. (1992)  {\em \apj, 396}, 178-194.

\refs Fisher J., Schr\"oder K.-P., and Smith R. C. (2005) {\em \mnras,
  361}, 495-503.

\refs Fisher R.T. (2004) {\em \apj, 600}, 769-780.

\refs Garc{\'{\i}}a B. and Mermilliod J. C. (2001) {\em \aap, 368},
122-136.

\refs Gawryszczak A. J., Goodwin S. P., Burkert A., and R\'o\.zyczka
M. (2006)  {\em \aap, submitted}.

\refs Gizis J. E., Reid I. N., Knapp G. R., Liebert J., Kirkpatrick
J. D., Koerner D. W., and Burgasser A. J. (2003)  {\em \aj, 125},
3302-3310.

\refs G\'omez L., Rodriguez L. F., Loinard L., Lizano S., Poveda
A., and Allen C. (2006) {\em \apj, in press} (astro-ph/0509201).

\refs Goodman A. A., Benson P. J., Fuller G. A., and Myers P. C.  (1993)
      {\em \apj, 406}, 528-547.

\refs Goodman A.A.,  Barranco J. A., Wilner D. J., and Heyer
M. H. (1998) {\em \apj, 504}, 223-246.

\refs Goodwin S. P. and Kroupa P. (2005) {\em \aap, 439}, 565-569.

\refs Goodwin S. P. and Whitworth A. P. (2004) {\em \aap, 413},
  929-937.

\refs Goodwin S. P., Ward-Thompson D., and Whitworth A. P. (2002) {\em
  \mnras, 330}, 769-771. 

\refs Goodwin S. P., Whitworth A. P., and Ward-Thompson D. (2004a) {\em
 \aap, 414}, 633-650.

\refs Goodwin S. P., Whitworth A. P., and Ward-Thompson D. (2004b) {\em
 \aap, 423}, 169-182.

\refs Goodwin S. P., Hubber D. A., Moraux E., and Whitworth
A. P. (2005)  {\em  Astron.Nachricten., 326}, 1040-1043.

\refs Hachisu I. and Eriguchi Y. (1984) {\em \aap, 140}, 259-264.

\refs Hachisu I. and Eriguchi Y. (1985) {\em \aap, 143}, 355-364.

\refs Haisch K. E. Jr., Greene T. P., Barsony M., and Stahler
S. W. (2004)  {\em \aj, 127}, 1747-1754.

\refs Heggie D. C. (1975) {\em  \mnras, 173}, 729-787.

\refs Heggie D. C., Hut, P., and McMillan S. L. W. (1996) {\em
      \apj, 467}, 359-369.

\refs Hennebelle P., Whitworth A. P., Gladwin P. P., and Andr\'e
P. (2003)  {\em \mnras,  340}, 870-882.

\refs Hennebelle P., Whitworth A. P., Cha S.-H., and Goodwin
S. P. (2004)  {\em \mnras, 348}, 687-701.

\refs Hennebelle P., Whitworth A. P., and Goodwin S. P. (2006) {\em
  \aap, in press}.

\refs Hills J. G. (1975) {\em \aj, 80}, 809-825.

\refs Hosking J. G. and Whitworth A. P. (2004a) {\em \mnras, 347}, 994-1000.

\refs Hosking J. G. and Whitworth A. P. (2004b) {\em \mnras, 347}, 1001-1010.

\refs Hubber D. A. and Whitworth A. P. (2005) {\em \aap, 437}, 113-125.

\refs Hubber D. A., Goodwin S. P., and Whitworth A. P. (2006) {\em
 \aap, in press}.

\refs Inutsuka S.-I. and Miyama S. M. (1992) {\em \apj, 388}, 392-399.

\refs Jappsen A-K. and Klessen R. S. (2004)  {\em \aap, 423}, 1-12.

\refs Jijina J., Myers P. C., and Adams F. C. (1999) {\em \apjs, 125}, 161-236.

\refs Jones C. E., Basu S., and Dubinski J. (2001) {\em \apj, 551}, 387-393.

\refs Kitsionas S. and Whitworth A. P. (2002) {\em \mnras, 330}, 129-136.

\refs Klessen R. S. and Burkert A. (2000) {\em \apjs, 128}, 287-319.

\refs Klessen R. S., Ballesteros-Paredes J., V\'azquez-Semadeni E., and
Dur\'an-Rojas (2005) {\em \apj, 620}, 786-794.

\refs K\"ohler R., and Leinert C. (1998) {\em \aap, 331}, 977-988.

\refs Kouwenhoven M. B. N., Brown A. G. A., Zinnecker H., Kaper L., 
and Portegies Zwart S. F. (2005) {\em \aap, 430}, 137-154.

\refs Kroupa P. (1995a) {\em \mnras, 277},  1491-1506.

\refs Kroupa P. (1995b) {\em \mnras, 277},  1507-1521.

\refs Kroupa P and Bouvier J. (2003a) {\em \mnras, 346}, 343-353.

\refs Kroupa P and Bouvier J. (2003b) {\em \mnras, 346}, 369-380

\refs Kroupa P. and Burkert A. (2001) {\em \apj, 555}, 945-949.

\refs Kroupa P., Tout C. A., and Gilmore G. (1993) {\em \mnras, 262}, 545-587.

\refs Kroupa P., Petr M. G., and McCaughrean M. J. (1999) {\em
  NewAstron., 4}, 495-520.

\refs Kroupa P., Aarseth S. J., and Hurley  J. (2001)  {\em
\mnras, 321}, 699-712.

\refs Kroupa P., Bouvier J., Duch{\^e}ne G., and Moraux E. (2003)
{\em \mnras, 346}, 354-368.

\refs Lada C. J. (2006) {\em \apj, submitted}.

\refs Lada C. J. and Lada E. A. (2003) {\em \araa, 41}, 57-115.

\refs Larson R. B. (1969) {\em \mnras, 145}, 297-308.

\refs Larson R. B. (1981) {\em \mnras, 194}, 809-826.

\refs Larson R. B. (2002) {\em \mnras, 332}, 155-164.

\refs Larson R. B. (2003) {\em Rep.Prog.Phys., 66}, 1651-1697.

\refs Leinert  C., Zinnecker  H., Weitzel  N., Christou  J., Ridgway 
S. T., Jameson  R., Haas  M., and Lenzen R. (1993) {\em \aap, 278},
129-149.

\refs Luhman K. L. (2004) {\em \apj, 617}, 1216-1232.

\refs Machida M. N., Matsumoto T., Tomisaka K., and Hanawa T. (2005a)
 {\em \mnras, 362}, 369-381.

\refs Machida M. N., Matsumoto T., Hanawa T., and Tomisaka K. (2005b)
 {\em \mnras, 362}, 382-402.

\refs Mart{\'{\i}}n E. L., Navascu{\' e}s D. B. y., Baraffe I., Bouy
        H., and Dahm S. (2003) {\em \apj, 594}, 525-532.

\refs Masunaga H. and Inutsuka S.(2000)  {\em \apj, 531}, 350-365.

\refs Masunaga H., Miyama S. M., and Inutsuka S. (1998) 
{\em \apj, 495}, 346-369.

\refs Mathieu R. D. (1994)  {\em \araa, 32}, 465-530.

\refs Maxted P. F. L. and Jeffries R. D. (2005) {\em \mnras, 362}, L45-L49.

\refs Mayor M, Duquennoy A., Halbwachs J.-L., and Mermilliod J.-C. 
(1992) In {\em Complementary Approaches to Double and Multiple Star 
Research} (H. A. McAlister and  W. I. Hartkopf, eds.), pp. 73-81. 
ASP Conf. Series, San Francisco.

\refs Mazeh T., Goldberg D., Duquennoy A., and Mayor M. (1992) 
 {\em \apj, 401}, 265-268.

\refs Miyama S. M. (1992)  {\em Pub.Astron.Soc.Jap., 44}, 193-202.

\refs Miyama S. M., Hayashi C., and Narita S. (1984)  
 {\em \apj,  279}, 621-632. 

\refs Motte F., Andr\'e P., and Neri R. (1998) {\em \aap, 336}, 150-172.

\refs Motte F., Andr\'e P., Ward-Thompson D., and Bontemps S. (2001)
      {\em \aap, 372}, L41-L44.

\refs Muench A. A., Lada E. A., Lada C. J., and Alves J. (2002) {\em
  \apj, 573}, 366-393.

\refs Myers P. C. (1983) {\em \apj, 270}, 105-118.

\refs Myers P. C. and Benson P. J. (1983) {\em \apj, 266}, 309-320.

\refs Myhill E. A. and Kaula W. M. (1992) {\em \apj, 386}, 578-586.

\refs Nelson R. P. and Papaloizou J. C. B. (1993) {\em \mnras, 265,} 905-920.

\refs Ochi Y., Sugimoto K., and Hanawa T. (2005) {\em \apj, 623}, 922-939.

\refs Padoan P. and Nordlund \AA (2002)  {\em \apj, 576}, 870-879.

\refs Padoan P. and Nordlund \AA (2004) {\em \apj, 617}, 559-564.

\refs Patience J., Ghez A. M., Reid I. N., and Matthews K. (2002) {\em
  \aj, 123}, 1570-1602.

\refs Pfalzner S. (2004) {\em \apj, 602}, 356-362.

\refs Pfalzner S., Vogel P. Scharw\"achter J., and Olczak C. (2005)
      {\em \aap, 437}, 967-976.

\refs  Pinfield D. J., Dobbie P. D., Jameson R. F., Steele I. A.,
Jones, H. R. A., and Katsiyannis A. C. (2003)  {\em
  \mnras, 342}, 1241-1259. 

\refs Preibisch T., Balega Y., Hofmann K., Weigelt G., and 
Zinnecker H. (1999) {\em NewAstron., 4}, 531-542.

\refs Price D. J. and Monaghan J. J. (2004a) {\em \mnras 348},
123-138.

\refs Price D. J. and Monaghan J. J. (2004b) {\em \mnras 348}, 139-152.

\refs Quist C. F. and Lindegren L. (2000)  {\em \aap, 361}, 770-780.

\refs Reid N. and Gizis J. E. (1997) {\em \aj 113}, 2246-2269.

\refs Reipurth B. (2000) {\em \aj, 120}, 3177-3191.

\refs Reipurth B. and Clarke C. J. (2001)  {\em \aj, 122}, 432-439.

\refs Reipurth B. and Zinnecker H. (1993)  {\em \aap, 278}, 81-108.

\refs Saigo K. and Tomisaka K. (2006) {\em \apj, submitted}.

\refs Schnee S. L., Ridge N. A., Goodman A. A., and Li J. G. (2005) 
{\em \apj, 634}, 442-450. 

\refs Shu F. H., Adams F. C., and Lizano S. (1987) {\em \araa, 25}, 23-81.

\refs Sigalotti L. D. G. and Klapp J. (1997) {\em \apj, 474}, 710-718.

\refs Sigalotti L. D. G. and Klapp J. (2000) {\em \apj, 531}, 1037-1052.

\refs S\"oderhjelm S. (2000)  {\em Astron.Nachricten., 321}, 165-170.

\refs Solomon P. M., Rivolo A. R., Barrett J., and Yahil A. (1987)
      {\em \apj, 319}, 730-741.

\refs Stamatellos D., Whitworth A. P., Boyd D. F. A., and Goodwin
S. P. (2005) {\em \aap, 439}, 159-169.

\refs Sterzik M. F. and Durisen R. H. (1998)  {\em \aap, 339}, 95-112.

\refs Sterzik M. F. and Durisen R. H. (2003)  {\em \aap, 400}, 1031-1042.

\refs Sterzik M. F., Durisen R. H., and Zinnecker, H. (2003) {\em \aap,
  411}, 91-97.

\refs Testi L. and Sargent A. (1998) {\em \apj, 508}, L91-L94.

\refs Thies I., Kroupa P., and Theis C. (2005) {\em \mnras, 364}, 961-970.

\refs Tohline J. E. (1981)  {\em \apj, 248}, 717-726.

\refs Tohline J. E. (1982) {\em Fund.Cosmic.Phys, 8}, 1-81.

\refs Tohline J. E. (2002) {\em \araa, 40}, 349-385.

\refs Tokovinin A. A. and Smekhov M. G. (2002) {\em \aap, 382}, 118-123.

\refs Truelove J. K., Klein R. I., McKee C. F. Holliman J. H. II,
Howell L. H., Grrenough J. A., and Woods D. T. (1997) {\em \apj, 495}, 821-852.

\refs Tsuribe T. and Inutsuka S.-I. (1999a) {\em \apj, 526}, 307-313.

\refs Tsuribe T. and Inutsuka S.-I. (1999b) {\em \apj, 523}, L155-L158.

\refs Umbreit S., Burkert A., Henning T., Mikkola S., and Spurzem
R. (2005)  {\em \apj, 623}, 940-951.

\refs V\'azquez-Semadeni E., Ostriker E. C., Passot T., Gammie C., and
Stone J. (2000) {\em Protostars and Planets IV} (V. Mannings et al., 
eds.), pp. 3-28. Univ. of Arizona, Tuscon.

\refs Watkins S. J., Bhattal  A. S., Boffin, H. M. J., Francis N., and 
Whitworth A. P. (1998a) {\em \mnras, 300}, 1205-1213.

\refs Watkins S. J., Bhattal  A. S., Boffin, H. M. J., Francis N., and 
Whitworth A. P. (1998b) {\em \mnras, 300}, 1214-1224.

\refs White R. J. and Ghez A. M. (2001) {\em \apj, 556}, 265-295.

\refs Whitehouse S. C. and Bate M. R. (2004) {\em \mnras, 353}, 1078-1094.

\refs Whitworth A. P., Chapman S. J., Bhattal A. S., Disney
M. J., Pongracic H., and Turner J. A. (1995) {\em \mnras, 277}, 727-746.

\refs Woitas J., Leinert C., and K\"ohler R. (2001), {\em \aap,
376}, 982-996.

\end{document}